# A systematic empirical comparison
# of different approaches for normalizing
# citation impact indicators

Ludo Waltman and Nees Jan van Eck

Centre for Science and Technology Studies, Leiden University, The Netherlands
{waltmanlr, ecknjpvan}@cwts.leidenuniv.nl

We address the question how citation-based bibliometric indicators can best be normalized to ensure fair comparisons between publications from different scientific fields and different years. In a systematic large-scale empirical analysis, we compare a traditional normalization approach based on a field classification system with three source normalization approaches. We pay special attention to the selection of the publications included in the analysis. Publications in national scientific journals, popular scientific magazines, and trade magazines are not included. Unlike earlier studies, we use algorithmically constructed classification systems to evaluate the different normalization approaches. Our analysis shows that a source normalization approach based on the recently introduced idea of fractional citation counting does not perform well. Two other source normalization approaches generally outperform the classification-system-based normalization approach that we study. Our analysis therefore offers considerable support for the use of source-normalized bibliometric indicators.

## 1. Introduction

Citation-based bibliometric indicators have become a more and more popular tool for research assessment purposes. In practice, there often turns out to be a need to use these indicators not only for comparing researchers, research groups, departments, or journals active in the same scientific field or subfield but also for making comparisons across fields (Schubert & Braun, 1996). Performing between-field comparisons is a delicate issue. Each field has its own publication, citation, and authorship practices, making it difficult to ensure the fairness of between-field comparisons. In some fields, researchers tend to publish a lot, often as part of larger collaborative teams. In other fields, collaboration takes place only at relatively small scales, usually involving no more than a few researchers, and the average publication output per researcher is significantly lower. Also, in some fields, publications tend to have long reference lists, with many references to recent work. In other fields, reference lists may be much shorter, or they may point mainly to older work. In the latter fields, publications on average will receive only a relatively small number of citations, while in the former fields, the average number of citations per publication will be much larger.

In this paper, we address the question how citation-based bibliometric indicators can best be normalized to correct for differences in citation practices between scientific fields. Hence, we aim to find out how citation impact can be measured in a way that allows for the fairest between-field comparisons.

In recent years, a significant amount of attention has been paid to the problem of normalizing citation-based bibliometric indicators. Basically, two streams of research can be distinguished in the literature. One stream of research is concerned with normalization approaches that use a field classification system to correct for differences in citation practices between scientific fields. In these normalization approaches, each publication is assigned to one or more fields and the citation impact of a publication is normalized by comparing it with the field average. Research into classification-system-based normalization approaches started in the late 1980s and the early 1990s (e.g., Braun & Glänzel, 1990; Moed, De Bruin, & Van Leeuwen, 1995). Recent contributions to this line of research were made by, among others, Abramo, Cicero, and D'Angelo (2012), Crespo, Herranz, Li, and Ruiz-Castillo (2012), Crespo, Li, and Ruiz-Castillo (2012), Radicchi and Castellano (2012c), Radicchi, Fortunato, and Castellano (2008), and Van Eck, Waltman, Van Raan, Klautz, and Peul (in press).

The second stream of research studies normalization approaches that correct for differences in citation practices between fields based on the referencing behavior of citing publications or citing journals. These normalization approaches do not use a field classification system. The second stream of research was initiated by Zitt and Small (2008),[1] who introduced the audience factor, an interesting new indicator of the citation impact of scientific journals. Other contributions to this stream of research were made by Glänzel, Schubert, Thijs, and Debackere (2011), Leydesdorff and

---

[1] Some first suggestions in the direction of this second stream of research were already made by Zitt, Ramanana-Rahary, and Bassecoulard (2005).

Bornmann (2011), Leydesdorff and Opthof (2010), Leydesdorff, Zhou, and Bornmann (2013), Moed (2010), Waltman and Van Eck (in press), Waltman, Van Eck, Van Leeuwen, and Visser (2013), Zhou and Leydesdorff (2011), and Zitt (2010, 2011). Zitt and Small referred to their proposed normalization approach as 'fractional citation weighting' or 'citing-side normalization'. Alternative labels introduced by other authors include 'source normalization' (Moed, 2010), 'fractional counting of citations' (Leydesdorff & Opthof, 2010), and 'a priori normalization' (Glänzel et al., 2011). Following our earlier work (Waltman & Van Eck, in press; Waltman et al., 2013), we will use the term 'source normalization' in this paper.

Which normalization approach performs best is still an open issue. Systematic large-scale empirical comparisons of normalization approaches are scarce, and as we will see, such comparisons involve significant methodological challenges. Studies in which normalization approaches based on a field classification system are compared with source normalization approaches have been reported by Leydesdorff, Radicchi, Bornmann, Castellano, and De Nooy (in press) and Radicchi and Castellano (2012a). In these studies, classification-system-based normalization approaches were found to be more accurate than source normalization approaches. However, as we will point out later on in this paper, these studies have important methodological limitations. In an earlier paper, we have compared a classification-system-based normalization approach with a number of source normalization approaches (Waltman & Van Eck, in press). The comparison was performed in the context of assessing the citation impact of scientific journals, and the results seemed to be in favor of some of the source normalization approaches. However, because of the somewhat non-systematic character of the comparison, the results must be considered of a tentative nature.

Building on our earlier work (Waltman & Van Eck, in press), we present in this paper a systematic large-scale empirical comparison of normalization approaches. The comparison involves one normalization approach based on a field classification system and three source normalization approaches. In the classification-system-based normalization approach, publications are classified into fields based on the journal subject categories in the Web of Science bibliographic database. The source normalization approaches that we consider are based on the audience factor approach of Zitt and Small (2008), the fractional citation counting approach of Leydesdorff and Opthof (2010), and our own revised SNIP approach (Waltman et al., 2013).

Our methodology for comparing normalization approaches has three important features not present in earlier work by other authors. First, rather than simply including all publications available in a bibliographic database in a given time period, we exclude as much as possible publications that could distort the analysis, such as publications in national scientific journals, popular scientific magazines, and trade magazines. Second, in the evaluation of the classification-system-based normalization approach, we use field classification systems that are different from the classification system used in the implementation of the normalization approach. In this way, we ensure that our results do not suffer from a bias that favors classification-system-based normalization approaches over source normalization approaches. Third, we compare normalization approaches at different levels of granularity, for instance both at the level of broad scientific disciplines and at the level of smaller scientific subfields. As we will see, some normalization approaches perform well at one level but not so well at another level.

To compare the different normalization approaches, our methodology uses a number of algorithmically constructed field classification systems. In these classification systems, publications are assigned to fields based on citation patterns. The classification systems are constructed using a methodology that we have introduced in an earlier paper (Waltman & Van Eck, 2012). Some other elements that we use in our methodology for comparing normalization approaches have been taken from the work of Crespo, Herranz, et al. (2012) and Crespo, Li, et al. (2012).

The rest of this paper is organized as follows. In Section 2, we discuss the data that we use in our analysis. In Section 3, we introduce the normalization approaches that we study. We present the results of our analysis in Section 4, and we summarize our conclusions in Section 5. The paper has three appendices. In Appendix A, we discuss the approach that we take to select core journals in the Web of Science database. In Appendix B, we discuss our methodology for algorithmically constructing field classification systems. Finally, in Appendix C, we report some more detailed results of our analysis.

## 2. Data

Our analysis is based on data from the Web of Science (WoS) bibliographic database. We use the Science Citation Index Expanded, the Social Sciences Citation

Index, and the Arts & Humanities Citation Index. Conference and book citation indices are not used. The data that we work with is from the period 2003–2011.

The WoS database is continuously expanding (Michels & Schmoch, 2012). Nowadays, the database contains a significant number of special types of sources, such as scientific journals with a strong national or regional orientation, trade magazines (e.g., *Genetic Engineering & Biotechnology News*, *Naval Architect*, and *Professional Engineering*), business magazines (e.g., *Forbes* and *Fortune*), and popular scientific magazines (e.g., *American Scientist*, *New Scientist*, and *Scientific American*). As we have argued in an earlier paper (Waltman & Van Eck, 2012), a normalization for differences in citation practices between scientific fields may be distorted by the presence of these special types of sources in one's database. For this reason, we do not simply include all WoS-indexed publications in our analysis. Instead, we include only publications from selected sources, which we refer to as WoS core journals. In this way, we intend to restrict our analysis to the international scientific literature covered by the WoS database. The details of our procedure for selecting publications in WoS core journals are discussed in Appendix A. Of the 9.79 million WoS-indexed publications of the document types *article* and *review* in the period 2003–2011, there are 8.20 million that are included in our analysis.

In the rest of this paper, the term 'publication' always refers to our selected publications in WoS core journals. Also, when we use the term 'citation' or 'reference', both the citing and the cited publication are assumed to belong to our set of selected publications in WoS core journals. Hence, citations originating from non-selected publications or references pointing to non-selected publications play no role in our analysis.

The analysis that we perform focuses on calculating the citation impact of publications from the period 2007–2010. There are 3.86 million publications in this period. For each publication, citations are counted until the end of 2011. The total number of citations equals 26.22 million. Notice that the length of the time window within which citations are counted is relatively short (i.e., between two and five years). When citation-based bibliometric indicators are used for research assessment purposes, it is common to have such short citation windows.

We use four different field classification systems in our analysis. One is the well-known system based on the WoS journal subject categories. In this system, a publication can belong to multiple research areas. The other three classification systems have been constructed algorithmically based on citation relations between publications. These classification systems, referred to as classification systems A, B, and C, differ from each other in their level of granularity. Classification system A is the least detailed system and consists of only 21 research areas. Classification system C, which includes 1,334 research areas, is the most detailed system. In classification systems A, B, and C, a publication can belong to only one research area. We refer to Appendix B for a discussion of the methodology that we have used for constructing classification systems A, B, and C. The methodology is largely based on an earlier paper (Waltman & Van Eck, 2012).

Table 1 provides some summary statistics for each of our four field classification systems. These statistics relate to the period 2007–2010. As mentioned above, our analysis focuses on publications from this period. Notice that in the WoS subject categories classification system the smallest research area ('Architecture') consists of only 94 publications. This is a consequence of the exclusion of publications in non-core journals. In fact, the total number of WoS subject categories in the period 2007–2010 is 250, but there are 15 categories (all in the arts and humanities) that do not have any core journal. This explains why there are only 235 research areas in the WoS subject categories classification system. In the other three classification systems, the overall number of publications is 3.82 million. This is about 1% less than the above-mentioned 3.86 million publications in the period 2007–2010. The reason for this small discrepancy is explained in Appendix B.

Table 1. Summary statistics for each of the four field classification systems.

| | No. of areas | Number of publications per area (2007–2010) | | | |
| --- | --- | --- | --- | --- | --- |
| | | Mean | Median | Minimum | Maximum |
| WoS subject categories | 235 | 27,524 | 16,448 | 94 | 191,790 |
| Classification system A | 21 | 182,133 | 137,548 | 49,577 | 635,209 |
| Classification system B | 161 | 23,757 | 19,085 | 4,800 | 69,816 |
| Classification system C | 1,334 | 2,867 | 2,421 | 820 | 12,037 |

## 3. Normalization approaches

As already mentioned, we study four normalization approaches in this paper, one based on a field classification system and three based on the idea of source normalization. In addition to correcting for differences in citation practices between scientific fields, we also want our normalization approaches to correct for the age of a publication. Recall that our focus is on calculating the citation impact of publications from the period 2007–2010 based on citations counted until the end of 2011. This means that an older publication, for instance from 2007, has a longer citation window than a more recent publication, for instance from 2010. To be able to make fair comparisons between publications from different years, we therefore need a correction for the age of a publication.

We start by introducing our classification-system-based normalization approach. In this approach, we calculate for each publication a *normalized citation score* (NCS). The NCS value of a publication is given by

$$\text{NCS} = \frac{c}{e}, \tag{1}$$

where $c$ denotes the number of citations of the publication and $e$ denotes the average number of citations of all publications in the same field and in the same year. Interpreting $e$ as a publication's expected number of citations, the NCS value of a publication is simply given by the ratio of the actual and the expected number of citations of the publication. An NCS value above (below) one indicates that the number of citations of a publication is above (below) what would be expected based on the field and the year in which the publication appeared. Averaging the NCS values of a set of publications yields the *mean normalized citation score* indicator discussed in an earlier paper (Waltman, Van Eck, Van Leeuwen, Visser, & Van Raan, 2011; see also Lundberg, 2007).

To determine a publication's expected number of citations $e$ in (1), we need a field classification system. In practical applications of the classification-system-based normalization approach, the journal subject categories in the WoS database are often used for this purpose.[2] For this reason, we also use the WoS subject categories in this paper, despite their limitations (e.g., Van Eck et al., in press). Notice that a publication may belong to multiple subject categories. In that case, we calculate the expected number of citations of the publication as the harmonic average of the expected numbers of citations obtained for the different subject categories. We refer to Waltman et al. (2011) for a justification of this approach.

---

[2] For instance, at our own institute, we perform large numbers of bibliometric studies in which we use the WoS subject categories to determine the expected number of citations of publications. In a similar way, the WoS subject categories are used in InCites, a web-based bibliometric analysis tool produced by Thomson Reuters.

We now turn to the three source normalization approaches that we study. In these approaches, a *source normalized citation score* (SNCS) is calculated for each publication. Since we have three source normalization approaches, we distinguish between the SNCS[1], the SNCS[2], and the SNCS[3] value of a publication. The general idea of the three source normalization approaches is to weight each citation received by a publication based on the referencing behavior of the citing publication or the citing journal.[3] The three source normalization approaches differ from each other in the exact way in which the weight of a citation is determined.

An important concept in the case of all three source normalization approaches is the notion of an active reference (Zitt & Small, 2008). In our analysis, an active reference is defined as a reference that falls within a certain reference window and that points to a publication in a WoS core journal. For instance, in the case of a four-year reference window, the number of active references in a publication from 2008 equals the number of references in this publication that point to publications in WoS core journals in the period 2005–2008. References to sources not covered by the WoS database or to WoS-indexed publications in non-core journals do not count as active references.

The SNCS[1] value of a publication is calculated as

$$\mathrm{SNCS}^{(1)} = \sum_{i=1}^{c} \frac{1}{a_i}, \qquad (2)$$

where $a_i$ denotes the average number of active references in all publications that appeared in the same journal and in the same year as the publication from which the $i$th citation originates. The length of the reference window within which active references are counted equals the length of the citation window of the publication for which the SNCS[1] value is calculated. The following example illustrates the definition of $a_i$. Suppose that we want to calculate the SNCS[1] value of a publication from 2008, and suppose that the *i*th citation received by this publication originates from a citing publication from 2010. Since the publication for which the $\text{SNCS}^{(1)}$ value is calculated has a four-year citation window (i.e., 2008–2011), $a_i$ equals the average number of active references in all publications that appeared in the citing journal in 2010, where active references are counted within a four-year reference window (i.e., 2007–2010). The $\text{SNCS}^{(1)}$ approach is based on the idea of the audience factor of Zitt and Small (2008), although it applies this idea to an individual publication rather than an entire journal. Unlike the audience factor, the $\text{SNCS}^{(1)}$ approach uses multiple citing years.

The $\text{SNCS}^{(2)}$ approach is similar to the $\text{SNCS}^{(1)}$ approach, but instead of the average number of active references in a citing journal it looks at the number of active references in a citing publication. In mathematical terms,

$$\text{SNCS}^{(2)} = \sum_{i=1}^{c} \frac{1}{r_i} \qquad (3)$$

where $r_i$ denotes the number of active references in the publication from which the *i*th citation originates. Analogous to the $\text{SNCS}^{(1)}$ approach, the length of the reference window within which active references are counted equals the length of the citation window of the publication for which the $\text{SNCS}^{(2)}$ value is calculated. The $\text{SNCS}^{(2)}$ approach is based on the idea of fractional citation counting of Leydesdorff and Opthof (2010; see also Leydesdorff & Bornmann, 2011; Leydesdorff et al., in press; Leydesdorff et al., 2013; Zhou & Leydesdorff, 2011).[4] However, a difference with the fractional citation counting idea of Leydesdorff and Opthof is that instead of all references in a citing publication only active references are counted. This is a quite important difference. Counting all references rather than active references only disadvantages fields in which a relatively large share of the references point to older literature, to sources not covered by the WoS database, or to WoS-indexed publications in non-core journals.

---

[4] In a somewhat different context, the fractional citation counting idea was already suggested by Small and Sweeney (1985).

The SNCS[3] approach, the third source normalization approach that we consider, combines ideas of the SNCS[1] and SNCS[2] approaches. The SNCS[3] value of a publication equals

$$\text{SNCS}^{(3)} = \sum_{i=1}^{c} \frac{1}{p_i r_i},$$ (4)

where $r_i$ is defined in the same way as in the SNCS[2] approach and where $p_i$ denotes the proportion of publications with at least one active reference among all publications that appeared in the same journal and in the same year as the $i$th citing publication. Comparing (3) and (4), it can be seen that the SNCS[3] approach is identical to the SNCS[2] approach except that $p_i$ has been added to the calculation. By including $p_i$, the SNCS[3] value of a publication depends not only on the referencing behavior of citing publications (like the SNCS[2] value) but also on the referencing behavior of citing journals (like the SNCS[1] value). The rationale for including $p_i$ is that some fields have more publications without active references than others, which may distort the normalization implemented in the SNCS[2] approach. For a more extensive discussion of this issue, we refer to Waltman et al. (2013), who present a revised version of the SNIP indicator originally introduced by Moed (2010). The SNCS[3] approach is based on similar ideas as this revised SNIP indicator, although in the SNCS[3] approach these ideas are applied to individual publications while in the revised SNIP indicator they are applied to entire journals. Also, the SNCS[3] approach uses multiple citing years, while the revised SNIP indicator uses a single citing year.

## 4. Results

We split the discussion of the results of our analysis in two parts. In Subsection 4.1, we present results that were obtained by using the WoS journal subject categories to evaluate the normalization approaches introduced in the previous section. We then argue that this way of evaluating the different normalization approaches is likely to produce biased results. In Subsection 4.2, we evaluate the normalization approaches using our algorithmically constructed classification systems A, B, and C instead of the WoS subject categories. We argue that this yields a fairer comparison of the different normalization approaches.

## 4.1. Results based on the Web of Science journal subject categories

Before presenting our results, we need to discuss how publications belonging to multiple WoS subject categories were handled. In the approach that we have taken, each publication is fully assigned to each of the subject categories to which it belongs. No fractionalization is applied. This means that some publications occur multiple times in the analysis, once for each of the subject categories to which they belong. Because of this, the total number of publications in the analysis is 6.47 million. The average number of subject categories per publication is 1.68.

Table 2 reports for each year in the period 2007–2010 the average normalized citation score of all publications from that year, where normalized citation scores have been calculated using each of the four normalization approaches introduced in the previous section. The average citation score (CS) without normalization is reported as well. As expected, unnormalized citation scores display a decreasing trend over time. This can be explained by the lack of a correction for the age of publications. Table 2 also lists the number of publications per year. Notice that each year the number of publications is 3% to 5% larger than the year before.

Table 2. Average normalized citation score per year calculated using four normalization approaches and the unnormalized CS approach. The citation scores are based on the 6.47 million publications included in the WoS journal subject categories classification system.

|  | 2007 | 2008 | 2009 | 2010 |
|---|---|---|---|---|
| No. of publications | 1.51M | 1.59M | 1.66M | 1.71M |
| CS | 10.78 | 8.16 | 5.50 | 2.70 |
| NCS | 1.01 | 1.01 | 1.02 | 1.02 |
| SNCS[(1)] | 1.10 | 1.07 | 1.07 | 1.05 |
| SNCS[(2)] | 1.03 | 0.97 | 0.89 | 0.68 |
| SNCS[(3)] | 1.10 | 1.07 | 1.07 | 1.05 |

Based on Table 2, we make the following observations:

- Each year, the average NCS value is slightly above one. This is a consequence of the fact that publications belonging to multiple subject categories are counted multiple times. Average NCS values of exactly one would have been obtained if there had been no publications that belong to more than one subject category.

- The average SNCS$^{(2)}$ value decreases considerably over time. The value in 2010 is more than 30% lower than the value in 2007. This shows that the SNCS$^{(2)}$ approach fails to properly correct for the age of a publication. Recent publications have a significant disadvantage compared with older ones. This is caused by the fact that in the SNCS$^{(2)}$ approach publications without active references give no 'credits' to earlier publications (see also Waltman & Van Eck, in press; Waltman et al., 2013). In this way, the balance between publications that provide credits and publications that receive credits is distorted. This problem is most serious for recent publications. In the case of recent publications, the citation and reference windows used in the calculation of SNCS$^{(2)}$ values are relatively short, and the shorter the length of the reference window within which active references are counted, the larger the number of publications without active references.

- The SNCS$^{(1)}$ and SNCS$^{(3)}$ approaches yield the same average values per year. These values are between 5% and 10% above one (see also Waltman & Van Eck, in press), with a small decreasing trend over time. Average SNCS$^{(1)}$ and SNCS$^{(3)}$ values very close to one would have been obtained if there had been no increase in the yearly number of publications (for more details, see Waltman & Van Eck, 2010; Waltman et al., 2013). The sensitivity of source normalization approaches to the growth rate of the scientific literature was already pointed out by Zitt and Small (2008).

Table 2 provides some insight into the degree to which the different normalization approaches succeed in correcting for the age of publications. However, the table does not show to what extent each of the normalization approaches manages to correct for differences in citation practices between scientific fields. This raises the question when exactly we can say that differences in citation practices between fields have been corrected for. With respect to this question, we follow a number of recent papers (Crespo, Herranz, et al., 2012; Crespo, Li, et al., 2012; Radicchi & Castellano, 2012a, 2012c; Radicchi et al., 2008). In line with these papers, we say that the degree to which differences in citation practices between fields have been corrected for is indicated by the degree to which the normalized citation distributions of different fields coincide with each other. Differences in citation practices between fields have been perfectly corrected for if, after normalization, each field is characterized by exactly the same citation distribution. Notice that correcting for the age of publications can be defined in an analogous way. We therefore say that publication age has been corrected for if different publication years are characterized by the same normalized citation distribution.

The next question is how the similarity of citation distributions can best be assessed. To address this question, we follow an approach that was recently introduced by Crespo, Herranz, et al. (2012) and Crespo, Li, et al. (2012). For each of the four normalization approaches that we study, we take the following steps:

1. Calculate each publication's normalized citation score.

2. For each combination of a publication year and a subject category, assign publications to quantile intervals based on their normalized citation score. We work with 100 quantile (or percentile) intervals. Publications are sorted in ascending order of their normalized citation score, and the first 1% of the publications are assigned to the first quantile interval, the next 1% of the publications are assigned to the second quantile interval, and so on.

3. For each combination of a publication year, a subject category, and a quantile interval, calculate the number of publications and the average normalized citation score per publication. We use $n(q, i, j)$ and $\mu(q, i, j)$ to denote, respectively, the number of publications and the average normalized citation score for publication year $i$, subject category $j$, and quantile interval $q$.

4. For each quantile interval, determine the degree to which publication age and differences in citation practices between fields have been corrected for. To do so, we calculate for each quantile interval $q$ the inequality index $I(q)$ defined as

$$I(q) = \frac{1}{n(q)} \sum_{i=2007}^{2010} \sum_{j=1}^{m} n(q,i,j) \frac{\mu(q,i,j)}{\mu(q)} \log\left( \frac{\mu(q,i,j)}{\mu(q)} \right), \qquad (5)$$

where $m$ denotes the number of subject categories and where $n(q)$ and $\mu(q)$ are given by, respectively,

$$n(q) = \sum_{i=2007}^{2010} \sum_{j=1}^{m} n(q,i,j) \qquad (6)$$

and

$$\mu(q) = \frac{1}{n(q)} \sum_{i=2007}^{2010} \sum_{j=1}^{m} n(q,i,j) \mu(q,i,j).$$ (7)

Hence, $n(q)$ denotes the number of publications in quantile interval $q$ aggregated over all publication years and subject categories, and $\mu(q)$ denotes the average normalized citation score of these publications. The inequality index $I(q)$ in (5) is known as the Theil index. We refer to Crespo, Li, et al. (2012) for a justification for the use of this index. The index takes non-negative values. The closer the value of the index is to zero, the better the correction for publication age and field differences. In the calculation of $I(q)$ in (5), we use natural logarithms and we define 0 log(0) = 0. Notice that $I(q)$ is not defined if $\mu(q) = 0$.

We perform the above steps for each of our four normalization approaches. Moreover, for the purpose of comparison, we perform the same steps also for citation scores without normalization.

The results of the above calculations are presented in Figure 1. For each of our four normalization approaches, the figure shows the value of $I(q)$ for each of the 100 quantile intervals. For comparison, $I(q)$ values calculated based on unnormalized citation scores are displayed as well. Notice that the vertical axis in Figure 1 has a logarithmic scale.

As expected, Figure 1 shows that all four normalization approaches yield better results than the approach based on unnormalized citation scores. For all or almost all quantile intervals, the latter approach, referred to as the CS approach in Figure 1, yields the highest $I(q)$ values. It can further be seen that the NCS approach significantly outperforms all three SNCS approaches. Hence, in line with recent studies by Leydesdorff et al. (in press) and Radicchi and Castellano (2012a), Figure 1 suggests that classification-system-based normalization is more accurate than source normalization. Comparing the different SNCS approaches, we see that the SNCS[(2)] approach is outperformed by the SNCS[(1)] and SNCS[(3)] approaches. Notice further that for all normalization approaches $I(q)$ values are highest for the lowest quantile intervals. These quantile intervals include many uncited and very lowly cited publications. The discreteness of citations therefore plays an important role in these quantile intervals, much more important than in the higher intervals. This discreteness explains the relatively high $I(q)$ values.

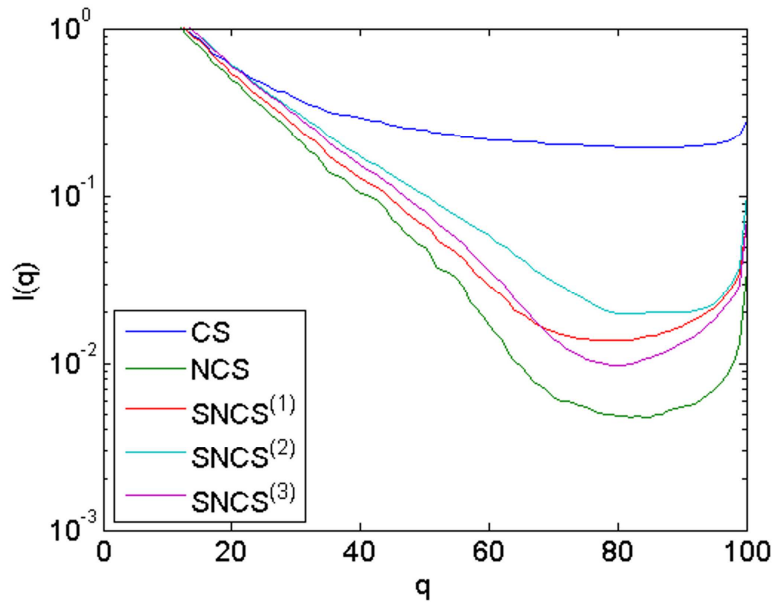

Figure 1. Inequality index $I(q)$ calculated for 100 quantile intervals $q$ and for four normalization approaches. Results calculated for the unnormalized CS approach are displayed as well. All results are based on the WoS journal subject categories classification system.

The above results may seem to provide clear evidence for preferring classification-system-based normalization over source normalization. However, there may be a bias in the results that causes the NCS approach to have an unfair advantage over the three SNCS approaches. The problem is that the WoS subject categories are used not only in the evaluation of the different normalization approaches but also in the implementation of one of these approaches, namely the NCS approach.[5] The standard used to evaluate the normalization approaches should be completely independent of the normalization approaches themselves, but for the NCS approach this is not the case. Because of this, the above results may be biased in favor of the NCS approach. In the next subsection, we therefore use our algorithmically constructed classification systems A, B, and C to evaluate the different normalization approaches in a fairer way.

Before proceeding to the next subsection, we note that the above-mentioned studies by Leydesdorff et al. (in press) and Radicchi and Castellano (2012a) suffer from the same problem as our above results. In these studies, the same classification system is used both in the implementation and in the evaluation of a classification-system-based normalization approach. This is likely to introduce a bias in favor of this normalization approach. This problem was first pointed out by Sirtes (2012) in a comment on Radicchi and Castellano's (2012a) study (for the rejoinder, see Radicchi & Castellano, 2012b).

### 4.2. Results based on classification systems A, B, and C

As we have argued above, for a fairer comparison of the four normalization approaches that we study, we need to evaluate the approaches using a classification system different from the WoS subject categories. We use the algorithmically constructed classification systems A, B, and C for this purpose. We note that we still use the WoS subject categories in the implementation of the NCS approach.

We believe that our algorithmically constructed classification systems can be expected to offer more accurate field definitions than the WoS subject categories. Contrary to the WoS subject categories, our algorithmically constructed classification systems are defined at the level of individual publications rather than at the journal level, making it easier to deal with publications in multidisciplinary journals. In addition, our algorithmically constructed classification systems provide a structure of science that is based not on the ideas of a small number of experts but instead on the collective actions of all publishing researchers, as reflected in their citation behavior.

---

classification system in which completely unrelated research areas are grouped together into fields. In other words, if the same classification system is used both in the implementation and in the evaluation of the NCS approach, the weaknesses of the classification system cannot be detected and the performance of the NCS approach is likely to be overestimated.

What we do not know is the level of granularity of an algorithmically constructed classification system that can be considered optimal from a normalization point of view. If fields are defined in a very broad way, they may not be sufficiently homogeneous in terms of citation behavior (e.g., Van Eck et al., in press). If fields are defined in a very narrow way, each publication is mainly compared with itself, creating a situation in which differences between publications are artificially reduced. Because the optimal level of granularity is unknown, we do not use a single algorithmically constructed classification system, but we use three of them. We think that the optimal level of granularity is probably somewhere in between classification systems B and C. Classification system A mainly serves to illustrate the effect of working with very broadly defined fields.

An obvious question is why we do not test implementations of the NCS approach in which classification systems A, B, and C are used instead of the WoS subject categories. The answer to this question is that the use of classification systems A, B, and C in both the implementation and the evaluation of the NCS approach would lead to the same problem as discussed in the previous subsection. The use of the same classification system in both the implementation and the evaluation of the NCS approach is likely to yield biased results in which the performance of the NCS approach is overestimated. For this reason, we cannot answer the question how the NCS approach implemented using classification system A, B, or C performs in comparison with our source normalization approaches.[6]

In classification systems A, B, and C, each publication belongs to only one research area. As explained in Section 2, the total number of publications included in the classification systems is 3.82 million. Based on these 3.82 million publications, Table 3 reports the average normalized citation score per year calculated using each of our four normalization approaches. The citation scores are very similar to the ones presented in Table 2. Like in Table 2, average NCS values are slightly above one. In the case of Table 3, this is due to the fact that of the 3.86 million publications in the period 2007–2010 a small proportion (about 1%) could not be included in classification systems A, B, and C (see Section 2).

---

[6] Some readers may nevertheless be interested to see the results obtained for the NCS approach implemented using classification systems A, B, and C. We have made these results available as supplementary material on www.ludowaltman.nl/normalization2/.

Table 3. Average normalized citation score per year calculated using four normalization approaches and the unnormalized CS approach. The citation scores are based on the 3.82 million publications included in classification systems A, B, and C.

|  | 2007 | 2008 | 2009 | 2010 |
|---|---|---|---|---|
| No. of publications | 0.90M | 0.94M | 0.98M | 1.01M |
| CS | 11.09 | 8.45 | 5.67 | 2.75 |
| NCS | 1.01 | 1.01 | 1.01 | 1.01 |
| SNCS[1] | 1.11 | 1.09 | 1.07 | 1.05 |
| SNCS[2] | 1.04 | 0.99 | 0.90 | 0.68 |
| SNCS[3] | 1.11 | 1.09 | 1.07 | 1.05 |

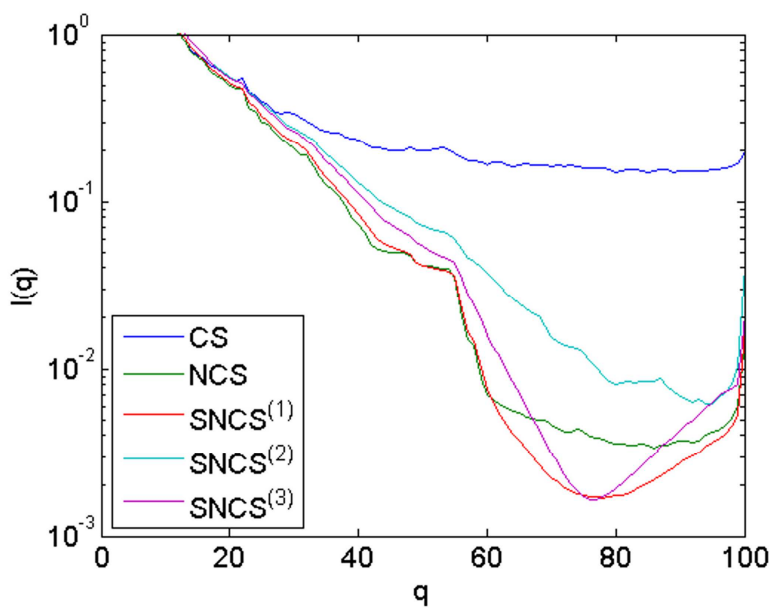

Figure 2. Inequality index $I(q)$ calculated for 100 quantile intervals $q$ and for four normalization approaches. Results calculated for the unnormalized CS approach are displayed as well. All results are based on classification system A.

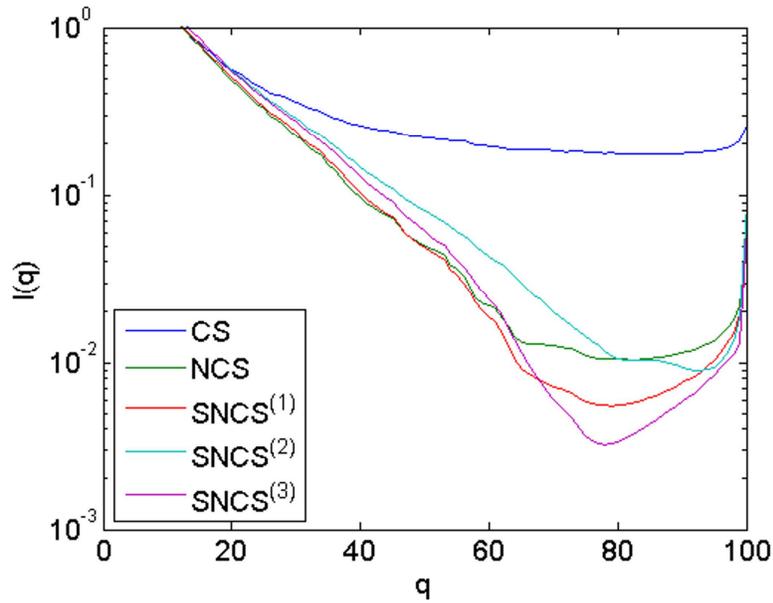

Figure 3. Inequality index $I(q)$ calculated for 100 quantile intervals $q$ and for four normalization approaches. Results calculated for the unnormalized CS approach are displayed as well. All results are based on classification system B.

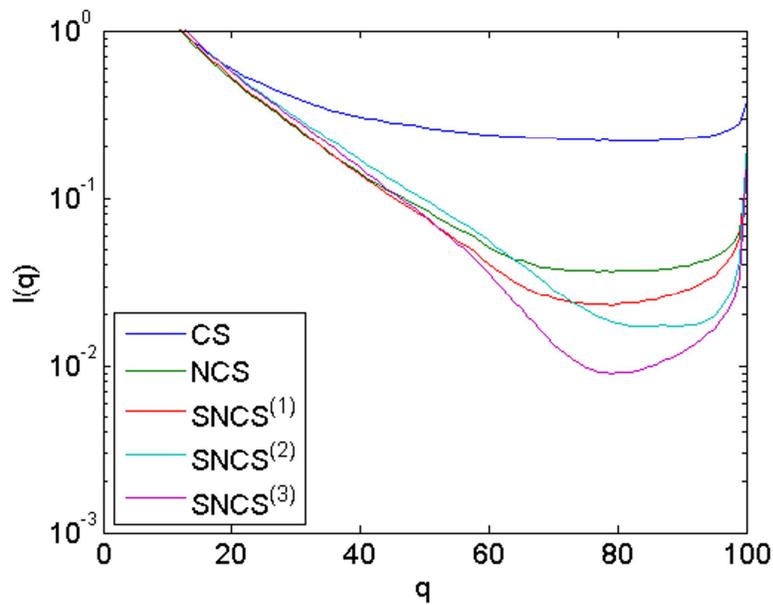

Figure 4. Inequality index $I(q)$ calculated for 100 quantile intervals $q$ and for four normalization approaches. Results calculated for the unnormalized CS approach are displayed as well. All results are based on classification system C.

We now examine the degree to which, after applying one of our four normalization approaches, different fields and different publication years are characterized by the same citation distribution. To assess the similarity of citation distributions, we take the same steps as described in the previous subsection, but with fields defined by research areas in our classification systems A, B, and C rather than by WoS subject categories. The results are shown in Figures 2, 3, and 4. Like in Figure 1, we use a logarithmic scale for the vertical axes.

The following observations can be made based on Figures 2, 3, and 4:

- Like in Figure 1, the CS approach, which does not involve any normalization, is outperformed by all four normalization approaches.

- The results presented in Figure 1 are indeed biased in favor of the NCS approach. Compared with Figure 1, the performance of the NCS approach in Figures 2, 3, and 4 is disappointing. In the case of classification systems B and C, the NCS approach is significantly outperformed by both the SNCS[1] and the SNCS[3] approach. In the case of classification system A, the NCS approach performs better, although it is still outperformed by the SNCS[1] approach.

- Like in Figure 1, the SNCS[2] approach is consistently outperformed by the SNCS[3] approach. In the case of classification systems A and B, the SNCS[2] approach is also outperformed by the SNCS[1] approach. It is clear that the disappointing performance of the SNCS[2] approach must at least partly be due to the failure of this approach to properly correct for publication age, as we have already seen in Tables 2 and 3.

- The SNCS[1] approach has a mixed performance. It performs very well in the case of classification system A, but not so well in the case of classification system C. The SNCS[3] approach, on the other hand, has a very good performance in the case of classification systems B and C, but this approach is outperformed by the SNCS[1] approach in the case of classification system A.

The overall conclusion based on Figures 2, 3, and 4 is that in order to obtain the most accurate normalized citation scores one should generally use a source normalization approach rather than a normalization approach based on the WoS subject categories classification system. However, consistent with our earlier work (Waltman & Van Eck, in press), it can be concluded that the SNCS[2] approach should not be used. Furthermore, the SNCS[(3)] approach appears to be preferable over the SNCS[(1)] approach. The excellent performance of the SNCS[(3)] approach in the case of classification system C (see Figure 4) suggests that this approach may be especially well suited for fine-grained analyses aimed for instance at comparing researchers or research groups active in closely related areas of research.

Some more detailed results are presented in Appendix C. In this appendix, we use a decomposition of citation inequality proposed by Crespo, Herranz, et al. (2012) and Crespo, Li, et al. (2012) to summarize in a single number the degree to which each of our normalization approaches has managed to correct for differences in citation practices between fields and differences in the age of publications.

## 5. Conclusions

In this paper, we have addressed the question how citation-based bibliometric indicators can best be normalized to ensure fair comparisons between publications from different scientific fields and different years. In a systematic large-scale empirical analysis, we have compared a normalization approach based on a field classification system with three source normalization approaches. In the classification-system-based normalization approach, we have used the WoS journal subject categories to classify publications into fields. The three source normalization approaches are inspired by the audience factor of Zitt and Small (2008), the idea of fractional citation counting of Leydesdorff and Opthof (2010), and our own revised SNIP indicator (Waltman et al., 2013).

Compared with earlier studies, our analysis offers three methodological innovations. Most importantly, we have distinguished between the use of a field classification system in the implementation and in the evaluation of a normalization approach. Following Sirtes (2012), we have argued that the classification system used in the evaluation of a normalization approach should be different from the one used in the implementation of the normalization approach. We have demonstrated empirically that the use of the same classification system in both the implementation and the evaluation of a normalization approach leads to significantly biased results. Building on our earlier work (Waltman & Van Eck, in press), another methodological innovation is the exclusion of special types of publications, for instance publications in national scientific journals, popular scientific magazines, and trade magazines. A third methodological innovation is the evaluation of normalization approaches at different levels of granularity. As we have shown, some normalization approaches perform better at one level than at another.

Based on our empirical results and in line with our earlier work (Waltman & Van Eck, in press), we advise against using source normalization approaches that follow the fractional citation counting idea of Leydesdorff and Opthof (2010). The fractional citation counting idea does not offer a completely satisfactory normalization (see also Waltman et al., 2013). In particular, we have shown that it fails to properly correct for the age of a publication.

The other two source normalization approaches that we have studied generally perform better than the classification-system-based normalization approach based on the WoS subject categories, especially at higher levels of granularity. It may be that other classification-system-based normalization approaches, for instance based on algorithmically constructed classification systems, have a better performance than subject-category-based normalization. However, any classification system can be expected to introduce certain biases in a normalization, simply because any organization of the scientific literature into a number of perfectly separated fields of science is artificial.[7] So consistent with our previous study (Waltman & Van Eck, in press), we recommend the use of a source normalization approach. Except at very low levels of granularity (e.g., comparisons between broad disciplines), the approach based on our revised SNIP indicator (Waltman et al., 2013) turns out to be more accurate than the approach based on the audience factor of Zitt and Small (2008). Of course, when using a source normalization approach, it should always be kept in mind that there are certain factors, such as the growth rate of the scientific literature, for which no correction is made.

---

[7] There are some other difficulties with the use of normalization approaches based on algorithmically constructed classification systems. Most importantly, it is unclear how to choose the most appropriate level of granularity for a classification system. Both the lowest possible level of granularity (i.e., all publications belonging to the same field) and the highest possible level of granularity (i.e., each publication belonging to its own field) are clearly of no use, so some intermediate level of granularity is needed. The difficulty is in determining which intermediate level of granularity provides the best normalization. Another difficulty is that algorithmically constructed classification systems (at the level of individual publications) need to be updated each time new publications become available. This updating is not only impractical, but depending on how it is done, it may also cause the field assignment of existing publications to be changed, thereby leading to unstable normalization results.

Some limitations of our analysis need to be mentioned as well. In particular, following a number of recent papers (Crespo, Herranz, et al., 2012; Crespo, Li, et al., 2012; Radicchi & Castellano, 2012a, 2012c; Radicchi et al., 2008), our analysis relies on a quite specific idea of what it means to correct for differences in citation practices between scientific fields. This is the idea that, after normalization, the citation distributions of different fields should coincide with each other as much as possible. There may well be alternative ways in which one can think of correcting for the field-dependent characteristics of citations. Furthermore, the algorithmically constructed classification systems that we have used to evaluate the different normalization approaches are subject to similar limitations as other classification systems of science. For instance, our classification systems artificially assume each publication to be related to exactly one research area. There is no room for multidisciplinary publications that belong to multiple research areas. Also, the choice of the three levels of granularity implemented in our classification systems clearly involves some arbitrariness.

Despite the limitations of our analysis, the conclusions that we have reached are in good agreement with three of our earlier papers. In one paper (Waltman et al., 2013), we have pointed out mathematically why a source normalization approach based on our revised SNIP indicator can be expected to be more accurate than a source normalization approach based on the fractional citation counting idea of Leydesdorff and Opthof (2010). In another paper (Waltman & Van Eck, in press), we have presented empirical results that support many of the findings of our present analysis. The analysis in our previous paper is less systematic than our present analysis, but it has the advantage that it offers various practical examples of the strengths and weaknesses of different normalization approaches. In a third paper (Van Eck et al., in press), we have shown, using a newly developed visualization methodology, that the use of the WoS subject categories for normalization purposes has serious problems. Many subject categories turn out not to be sufficiently homogeneous to serve as a solid base for normalization. Altogether, we hope that our series of papers will help to significantly increase the fairness of bibliometric research assessments, in particular of multidisciplinary assessments involving comparisons of citation impact between different fields of science.

## Acknowledgments

We would like to thank our colleagues at the Centre for Science and Technology Studies for their feedback on this research project. We are grateful to Loet Leydesdorff, Filippo Radicchi, and Javier Ruiz-Castillo for their comments on an earlier version of this paper and for helpful discussions.

## Appendix A: Selection of Web of Science core journals

In this appendix, we discuss the procedure that we have used for selecting core journals in the WoS database in the period 2003–2011. The procedure consists of three steps.

In step 1, we identify all publications in the WoS database that satisfy each of the following four criteria:

- The publication is of the document type *article* or *review*.
- The publication has at least one author (or group author).
- The publication is in English (Van Leeuwen, Moed, Tijssen, Visser, & Van Raan, 2001; Van Raan, Van Leeuwen, & Visser, 2011a, 2011b).

- The publication appeared in the period 1999–2011. (The reason for including publications not only from the period 2003–2011 but also from the period 1999–2002 will become clear in step 3 discussed below.)

In the rest of this appendix, the term 'publication' refers only to publications that satisfy the above four criteria.

In step 2, we aim to distinguish between international journals and journals with a strong focus on one or a few countries (see also Buela-Casal, Perakakis, Taylor, & Checa, 2006; Waltman & Van Eck, in press; Zitt & Bassecoulard, 1998; Zitt, Ramanana-Rahary, & Bassecoulard, 2003). We first determine for each publication the countries to which it belongs. This is done based on the addresses listed in a publication (including the reprint address). If a publication has multiple addresses in the same country, the country is counted only once. For instance, a publication with two Dutch addresses and one Belgian address is considered to belong half to the Netherlands and half to Belgium.

Based on the publication-country links, we determine for each journal a distribution over countries. We start by determining a distribution over countries for each individual publication in a journal. This is done by looking not only at the countries to which a publication itself belongs but also at the countries to which citing publications (if any) belong. Suppose for instance that we have a mixed Dutch-Belgian-French publication and that this publication is cited by a Dutch publication and by a mixed Dutch-Belgian-German publication. The publication then has a distribution that gives a weight of 5/9 to the Netherlands, a weight of 2/9 to Belgium, and a weight of 1/9 to France and Germany each. The distribution over countries for a journal as a whole is obtained simply by averaging the distributions of the individual publications in the journal. Each publication has equal weight, regardless of the number of times it has been cited. Publications for which no country data is available (e.g., because they do not list any addresses and because they have not been cited) are ignored.

To distinguish between international journals and journals with a strong focus on one or a few countries, we compare each journal's distribution over countries with the overall distribution over countries based on all publications identified in step 1. Like in an earlier paper (Waltman & Van Eck, in press), we use the Kullback-Leibler divergence for comparing the two distributions.[8] For a given journal $i$, the Kullback-Leibler divergence equals

$$d_i = \sum_j p_{ij} \ln \frac{p_{ij}}{q_j}, \tag{A1}$$

where $p_{ij}$ denotes the weight of country $j$ in journal $i$'s distribution over countries and $q_j$ denotes the weight of country $j$ in the overall distribution over countries. The higher the value of $d_i$, the stronger the focus of journal $i$ on one or a few countries.

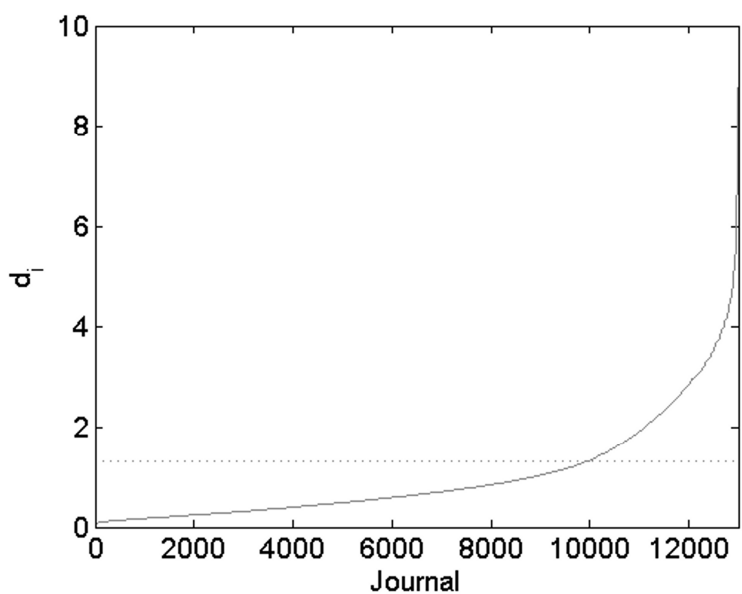

Figure A1. Distribution of journals' $d_i$ values. The horizontal line indicates the threshold of 1.3260.

Figure A1 shows the distribution of journals' $d_i$ values. The horizontal line, drawn at a $d_i$ value of 1.3260, indicates the threshold that we have chosen to distinguish between international and non-international journals. Journals with a $d_i$ value below this threshold are considered to be international. The remaining journals are considered not to be sufficiently internationally oriented. These journals are excluded as core journals. Choosing a threshold for $d_i$ necessarily involves some arbitrariness. In this case, we have chosen a threshold of 1.3260 because this is the highest threshold for which all journals that are fully focused on a single country only are excluded (see also Waltman & Van Eck, in press). Increasing the threshold would cause journals that are completely US oriented (i.e., all addresses of all publications in the journal itself and of all citing publications are in the US) to be considered international.

---

[8] An attractive property of the Kullback-Leibler divergence is that under certain assumptions it is insensitive to splits and mergers of countries. Suppose for instance that instead of treating the countries of the European Union (EU) as separate entities the EU is treated as one big entity. If it is assumed that journals have their EU activity distributed proportionally over the EU countries (i.e., proportionally to each country's overall weight), then journals' Kullback-Leibler values will remain unchanged.

This brings us to step 3 of our procedure for selecting core journals. The aim of this final step is to exclude journals that in terms of their referencing behavior are not sufficiently connected to the recent WoS-indexed scientific literature. This could be journals that simply do not cite very often, such as many trade magazines, business magazines, and popular scientific magazines, but this could also be journals, for instance in the arts and humanities and in some of the social sciences, that do cite quite a lot but mainly to sources not covered by the WoS database. The approach that we take to implement step 3 is similar to the journal selection approach taken in our revised SNIP indicator (Waltman et al., 2013).

In step 3, we look at the publications in a journal in the period 2003–2011, and for each publication we determine whether it contains at least one citation to a recent publication in a WoS core journal. A cited publication is considered to be recent if it appeared no more than four years before the citing publication. (This is why in step 1 discussed above we included publications not only from the period 2003–2011 but also from the period 1999–2002.) It is required that at least 50% of the publications in a journal cite at least one recent publication in a WoS core journal. Journals that do not meet this requirement are excluded as core journals. Notice that excluding a journal as a core journal may cause another journal that initially was just above the 50% threshold to fall below this threshold. In that case, the latter journal is excluded as well. In this way, the set of core journals is iteratively reduced until we are left with a stable set of journals that all meet the 50% requirement.

Some results of our three-step procedure for selecting core journals in the WoS database are summarized in Table A1. The table lists the initial number of publications (*article* and *review* document types only) and journals and the remaining number of publications and journals after each of the three steps.

Table A1. Results of the proposed three-step procedure for selecting core journals in the WoS database. The time period is 2003–2011.

|              | No. of publications | No. of journals |
| ------------ | ------------------- | --------------- |
| Initially    | 9.79M               | 13,263          |
| After step 1 | 9.31M               | 13,040          |
| After step 2 | 8.47M               | 9,980           |
| After step 3 | 8.20M               | 8,761           |

# Appendix B: Algorithmic construction of field classification systems

In this appendix, we discuss the methodology that we have used for algorithmically constructing three classification systems of WoS publications. Most of the methodology has been taken from an earlier paper (Waltman & Van Eck, 2012). We therefore refer to this paper for a more detailed discussion of the methodology.

Our aim was to construct three different classification systems of the 8,199,827 publications in WoS core journals in the period 2003–2011. A classification system assigns each publication to a research area. Publications can belong to one research area only. Unlike in our earlier work (Waltman & Van Eck, 2012), no hierarchy of research areas was created. The three classification systems differ from each other in their level of granularity. Classification system A consists of the smallest number of research areas and therefore offers the lowest level of detail. Classification system C includes the largest number of research areas and therefore provides the highest level of detail. Classification system B is in between the other two systems.

Our methodology for constructing a classification system starts by identifying all direct citation relations between the 8.20 million publications of interest. A total of 80.56 million direct citation relations were identified. The direction of a citation is ignored in our methodology, so citing and cited publications are treated symmetrically. Co-citation and bibliographic coupling relations play no role in our methodology. Initially, only publications belonging to the largest connected component of the direct citation network are included in a classification system. The largest connected component turned out to consist of 7.88 million publications.

To assign publications to research areas, first a normalization procedure is applied, as discussed in our earlier work (Waltman & Van Eck, 2012). This normalization procedure aims to ensure that research areas in different disciplines will be of about the same size, despite large differences in citation practices between disciplines. After the normalization procedure has been applied, a clustering technique is used to perform the assignment of publications to research areas. The clustering technique that we use (Waltman & Van Eck, 2012) is a variant of the well-known technique of modularity-based clustering (Newman, 2004; Newman & Girvan, 2004) and is also closely related to the technique that we have introduced as part of our unified approach to mapping and clustering of bibliometric networks (Waltman, Van Eck, & Noyons, 2010).

In the assignment of publications to research areas, there are two parameters for which a suitable value needs to be chosen. One is the so-called resolution parameter, which determines the number of research areas in a classification system. In general, the higher the value of this parameter, the larger the number of research areas and, consequently, the higher the level of detail offered by the classification system. The other parameter is the minimum number of publications per research area. For each of our three classification systems, Table B1 lists the values that we have chosen for the two parameters. Using these parameter values, the number of research areas that we obtained for classification systems A, B, and C is, respectively, 21, 161, and 1,334.

Table B1. Resolution parameter and minimum number of publications per research area for each of the three classification systems.

|  | Resolution | Min. no. of pub. per area |
|---|---|---|
| Classification system A | $1\times10^{-7}$ | 100,000 |
| Classification system B | $5\times10^{-7}$ | 10,000 |
| Classification system C | $5\times10^{-6}$ | 2,000 |

As mentioned above, initially only the 7.88 million publications belonging to the largest connected component of the direct citation network were included in the classification systems. In the final step of our methodology, an attempt is made to also include the remaining publications. This is done based on bibliographic coupling relations. Each publication is added to the research area with which it is most strongly bibliographically coupled. In this way, each of our classification systems in the end included 8,117,743 publications. This means that of the 8,199,827 publications in WoS core journals in the period 2003–2011, there are 82,084 publications (1.0%) that could not be included in our classification systems. These publications have no direct citation relations and no bibliographic coupling relations with publications that are included in the classification systems.

## Appendix C: Decomposition of citation inequality

One may wish to have a single number that summarizes the degree to which a normalization approach has managed to correct for differences in citation practices between scientific fields and differences in the age of publications. For this purpose, a decomposition of citation inequality was proposed by Crespo, Herranz, et al. (2012) and Crespo, Li, et al. (2012). In this appendix, we first briefly summarize the idea of this decomposition and we then present the results obtained based on classification systems A, B, and C.

Suppose we have $n$ publications belonging to a number of different fields and publication years. Let $c_i$ denote the normalized or unnormalized citation score of publication $i$. The overall citation inequality $I$ is defined as

$$I = \frac{1}{n} \sum_{i=1}^{n} \frac{c_i}{\mu} \log\left(\frac{c_i}{\mu}\right),$$  (C1)

where $\mu$ denotes the average citation score per publication. Hence, $\mu$ is given by

$$\mu = \frac{1}{n} \sum_{i=1}^{n} c_i .$$  (C2)

Crespo, Herranz, et al. (2012) and Crespo, Li, et al. (2012) point out that the overall citation inequality $I$ can be written as

$$I = W + S + IDCP ,$$  (C3)

where $W$ captures the citation inequality within each quantile interval for each combination of a field and a publication year and $S$ captures the inequality in the $\mu(q)$ values given by (7). *IDCP* captures the *inequality due to differences in citation practices*, that is, it captures the inequality in the $\mu(q, i, j)$ values for each quantile interval $q$. In fact, *IDCP* equals

$$IDCP = \frac{1}{n\mu} \sum_{q=1}^{100} n(q)\mu(q)I(q) \,, \tag{C4}$$

where $I(q)$, $n(q)$, and $\mu(q)$ are given by, respectively, (5), (6), and (7). In other words, $IDCP$ equals a weighted average of the inequality indices $I(q)$ of the different quantile intervals, where for each quantile interval the inequality index $I(q)$ is weighted by $n(q)\mu(q)$, which is the total citation score of the publications belonging to the quantile interval. The lower the value of $IDCP$, the better differences in citation practices between fields and differences in the age of publications have been corrected for.

Results obtained based on classification systems A, B, and C are reported in Tables C1, C2, and C3. For each classification system, values of $I$, $W$, $S$, and $IDCP$ have been calculated using each of the four normalization approaches that we study and also using the unnormalized CS approach. The $IDCP$ values reported in Tables C1, C2, and C3 confirm the observations made based on Figures 2, 3, and 4. Notice, however, that the differences between the $IDCP$ values of the different normalization approaches are relatively small. This is caused by the fact that the highest quantile intervals have a lot of weight in the $IDCP$ calculation in (C4). As can be seen in Figures 2, 3, and 4, in the highest quantile intervals, the differences between the normalization approaches are not so large.

Table C1. Values of $I$, $W$, $S$, and $IDCP$ calculated using four different normalization approaches and the unnormalized CS approach. The values are based on classification system A.

|              | $W$    | $S$    | $IDCP$  | $I$    |
|--------------|--------|--------|---------|--------|
| CS           | 0.0292 | 0.7066 | 0.1818  | 0.9176 |
| NCS          | 0.0221 | 0.6619 | 0.0237  | 0.7076 |
| SNCS$^{(1)}$ | 0.0291 | 0.6723 | 0.0235  | 0.7249 |
| SNCS$^{(2)}$ | 0.0370 | 0.6328 | 0.0379  | 0.7077 |
| SNCS$^{(3)}$ | 0.0337 | 0.6611 | 0.0266  | 0.7215 |

Table C2. Values of *I*, *W*, *S*, and *IDCP* calculated using four different normalization approaches and the unnormalized CS approach. The values are based on classification system B.

|  | *W* | *S* | *IDCP* | *I* |
|---|---|---|---|---|
| CS | 0.0247 | 0.6787 | 0.2142 | 0.9176 |
| NCS | 0.0194 | 0.6508 | 0.0374 | 0.7076 |
| SNCS[(1)] | 0.0253 | 0.6640 | 0.0355 | 0.7249 |
| SNCS[(2)] | 0.0326 | 0.6272 | 0.0479 | 0.7077 |
| SNCS[(3)] | 0.0300 | 0.6565 | 0.0349 | 0.7215 |

Table C3. Values of *I*, *W*, *S*, and *IDCP* calculated using four different normalization approaches and the unnormalized CS approach. The values are based on classification system C.

|  | *W* | *S* | *IDCP* | *I* |
|---|---|---|---|---|
| CS | 0.0164 | 0.6295 | 0.2717 | 0.9176 |
| NCS | 0.0135 | 0.6185 | 0.0756 | 0.7076 |
| SNCS[(1)] | 0.0169 | 0.6389 | 0.0690 | 0.7249 |
| SNCS[(2)] | 0.0217 | 0.6138 | 0.0722 | 0.7077 |
| SNCS[(3)] | 0.0202 | 0.6449 | 0.0563 | 0.7215 |